\documentclass[prl,twocolumn,superscriptaddress]{revtex4}
\usepackage{graphicx}
\begin{document}

\title{The de Haas-van Alphen effect across the metamagnetic transition in 
Sr$_3$Ru$_2$O$_7$}

\author{R.~A.~Borzi}
\affiliation{School of Physics and Astronomy, University of St. Andrews, St. Andrews KY16 9SS, UK.}
\author{S.~A.~Grigera}
\affiliation{School of Physics and Astronomy, University of St. Andrews, St. Andrews KY16 9SS, 
UK.}
\author{R.~S.~Perry}
\affiliation{Kyoto University International Innovation Center, Kyoto 606-8501, Japan.}
\affiliation{Department of Physics, Kyoto University, Kyoto 606-8502, Japan.}
\author{N.~Kikugawa}
\affiliation{School of Physics and Astronomy, University of St. Andrews, St. Andrews KY16 9SS, 
UK.}
\author{K.~Kitagawa}
\affiliation{Kyoto University International Innovation Center, Kyoto 606-8501, Japan.}
\author{Y.~Maeno}
\affiliation{Kyoto University International Innovation Center, Kyoto 606-8501, Japan.}
\affiliation{Department of Physics, Kyoto University, Kyoto 606-8502, Japan.}
\author{A.~P.~Mackenzie}
\affiliation{School of Physics and Astronomy, University of St. Andrews, St. Andrews KY16 9SS, 
UK.}

\begin{abstract}
We report a study of the de Haas-van Alphen (dHvA) effect on the itinerant 
metamagnet Sr$_3$Ru$_2$O$_7$.  Extremely high sample purity allows the observation of 
dHvA oscillations both above and below the metamagnetic transition field of 
7.9 T.  The quasiparticle masses are fairly large away from the transition, 
and are enhanced by up to an extra factor of three as the transition is 
approached, but the Fermi surface topography change is quite small.  The 
results are qualitatively consistent with a field-induced Stoner transition 
in which the mass enhancement is the result of critical fluctuations.\\

\noindent PACS numbers: 71.27.+a, 71.18.+y, 74.70.Pq

\end{abstract}
\maketitle

The applicability of Landau's Fermi liquid theory to strongly correlated metals such as oxides continues to be one of the major topics of interest in condensed matter physics \cite{[1]}.  Given the on-going debate about the nature of the metallic states in these systems, it is attractive to probe the border between Fermi liquid (FL) and non-Fermi liquid (NFL) states.   This border can be accessed in a controlled manner in `quantum critical' systems in which externally applied tuning parameters are used to depress the characteristic temperature of a second order phase transition~\cite{[2]} or critical end-point~\cite{[3]} to low temperatures.  Pressure~\cite{[4]} and chemical composition~\cite{[5]} are useful methods of tuning towards quantum criticality, but magnetic field enables a wide range of experiments on samples that need not contain the disorder that is inherent in most forms of chemical doping.  Magnetic field tuning of quantum phase transitions has been used in a variety of systems from insulators~\cite{[6]} to magnetic multilayers~\cite{[7]} and itinerant antiferromagnets~\cite{[8]}.  Recently, our interest has focused on its applicability to itinerant metamagnets, paramagnetic metals that show a non-linear rise in magnetisation at finite applied field.  In these systems, there is no change in symmetry between the low- and high-field states, so any phase transition must be first-order, and quantum criticality can only result from tuning of a critical end-point~\cite{[3]}.  

The material used as a model system in our work is the layered ruthenate metal Sr$_3$Ru$_2$O$_7$, which is an enhanced paramagnet with a metamagnetic transition whose characteristic field $B_c$ varies from approximately 5.1~T to 7.9~T as the applied field is rotated from the $ab$ plane to the $c$ axis~\cite{[9],[10]}.  In single crystals with residual resistivity ($\rho_{\rm res}$) of approximately 3~$\mu\Omega$cm, transport, dynamical susceptibility and thermodynamic measurements provided evidence for the existence of a quantum critical end-point for magnetic fields {\bf B} applied parallel to the $c$ axis~\cite{[10],[11]}.  Very near the metamagnetic field (specifically for $T < 0.8~{\rm K}$ and $B=B_c \pm 0.03~{\rm T}$) there are indications of transport properties that could not be a straightforward consequence of critical fluctuations~\cite{[3]}.   To investigate whether the anomalous region is intrinsic, a further generation of samples with even lower levels of disorder $(\rho_{\rm res} = 0.4~\mu\Omega{\rm cm})$ has been grown.  In these new samples, the anomalous behaviour is much more pronounced, and is seen in a larger range of fields near $B_c$, bounded at low temperatures by two first-order phase transitions at $B_c$  =  7.86 and 8.10~T~\cite{[12]}.  Outside this field range, however, the transport measurements indicate that the system still behaves as if there were a quantum critical point at $B_c$.  

There is a clear motivation to perform microscopic measurements of the behaviour of Landau quasiparticles above and below the metamagnetic transition in Sr$_3$Ru$_2$O$_7$, in order to see how they evolve on the approach to criticality.  This can be done via the de Haas - van Alphen effect, in which oscillations in the magnetisation result from quantization of cyclotron motion of charged quasiparticles in planes normal to {\bf B}.  The oscillations are periodic in $1/B$  with a frequency $F$ which gives the area $A$ of an extremal Fermi surface cross-section normal to {\bf B} via $A = 2\pi F/\hbar$.  The amplitude is strongly damped both thermally and by elastic impurity scattering, placing strong demands on material quality if a successful experiment is to be performed.  Analysis of the temperature dependence yields the quasiparticle effective mass $m$, which can be much larger than the bare electron mass in correlated electron systems of interest.

With the relatively low value of $B_c \approx 7.9~{\rm T}$ in Sr$_3$Ru$_2$O$_7$, extremely high sample purity is required for the oscillations to be resolved in both the low- and high-field states.  In this paper, we report a de Haas-van Alphen study in which we demonstrate that this is now possible on the new generation of samples.  Oscillations can be resolved in the field ranges 5~T $ < B < $ 7.5~T and 8.3~T $< B < $ 15~T with sufficient precision to enable studies of the field dependence of the quasiparticle mass and Fermi surface topography.  The picture that emerges is one of the metamagnetism being consistent with a Stoner-like transition of the itinerant system, accompanied by a strong enhancement of the quasiparticle mass as the transition is approached from either above or below.  The measurements therefore give direct microscopic support for the idea that if a `coarse grained' view is taken sufficiently far from the metamagnetic transition, the system's properties appear to be governed by the presence of a quantum critical point.

The crystals used in this project were grown in an image furnace in Kyoto \cite{[12]}.  The measurements were performed in St. Andrews, using a cryomagnetic system comprising a dilution refrigerator and superconducting magnets, and incorporating low temperature passive electronics.  The de Haas-van Alphen effect was observed in four samples, with the  field parallel to $c$.  The results shown here are from the samples showing the largest oscillatory amplitude and hence the best signal to noise ratio.  Heating due to eddy currents in the rotation stage restricted us to a low modulation field of 0.15 Oe, applied at 83.4 Hz through a superconducting modulation coil. For this reason, detection was at the first harmonic, so the data are proportional to $\chi (\omega)$~\cite{[14]}.  Voltage noise levels were typically less than 50~pV/$\sqrt{{\rm Hz}}$.

%
\begin{figure}
\includegraphics[angle=-90,width=\columnwidth]{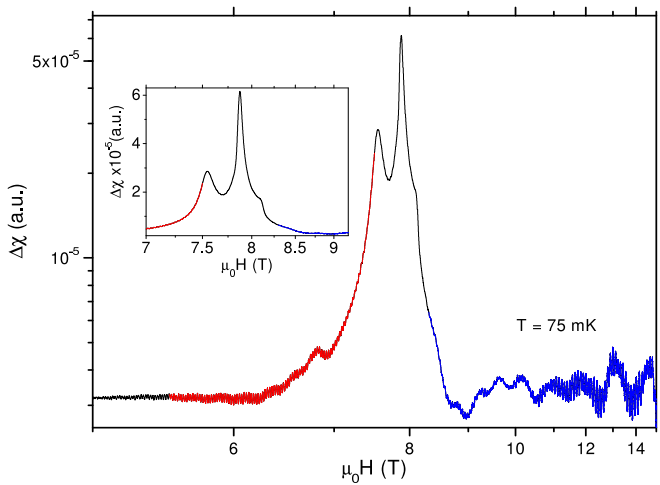}
\caption[]{(color) The real part of the differential magnetic susceptibility
through the metamagnetic transition in Sr$_3$Ru$_2$O$_7$ on a logarithmic scale.  The oscillations are seen below (red) and above (blue) the transition , but are almost indistinguishable in a linear scale (inset).}
\end{figure}

In Fig. 1 we show the real part of the susceptibility from 5~T to 15~T.  On the logarithmic scale used, the tiny oscillations are visible, but they are essentially impossible to resolve with the naked eye on a linear scale (inset).  The coloured regions denote the field ranges over which oscillations can be resolved after subtraction of the slowly varying background.  The first order magnetic transitions~\cite{[12]} discussed in the introduction correspond to the main peak and to the high-field shoulder.  There is no evidence of first order behaviour associated with the peak at 
7.55~T~.

%
\begin{figure}
\includegraphics[angle=-90,width=\columnwidth]{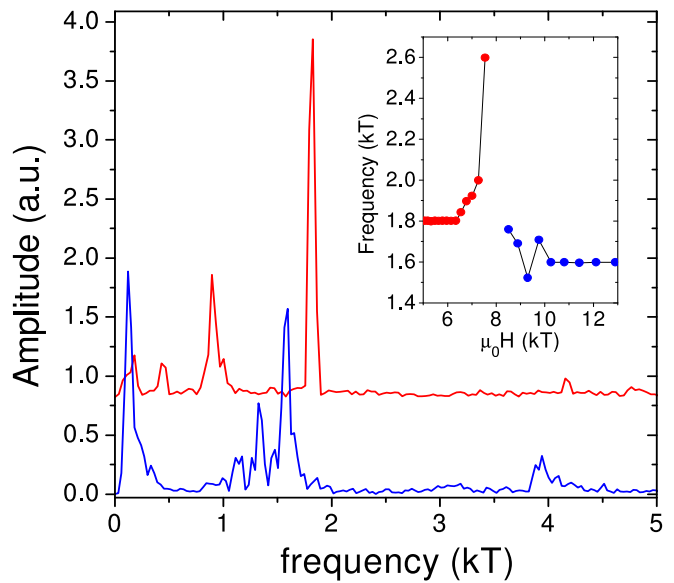}
\caption[]{(color) The Fourier transformation of the oscillatory data of Fig. 1.  The color coding corresponds to that of Fig. 1: red and blue for below and above the transition respectively.  The transformations were performed over the regions 5.5~T $< B <~ $6.5~T and 10~T $ < B <~ $15~T, where the frequency is field-independent (see inset).  The inset shows the field dependence of the main peak frequency below and above the transition(1.8 kT  and 1.6 kT respectively)}
\end{figure}

Fourier transformations of the oscillatory data above and below the metamagnetic field are shown in Fig.~2.  To begin the interpretation of these data, we refer back to the literature on heavy fermion systems.  Sr$_3$Ru$_2$O$_7$ is not the first material in which quantum oscillations have been observed both above and below a metamagnetic transition.  Notable examples include work on UPt$_3$ and CeRu$_2$Si$_2$ by the Cambridge and Osaka groups~\cite{[16],[17],[18],[19]}.  As discussed in those papers, it is important to distinguish between the `true' and observed dHvA frequencies in metamagnets or other systems with non-linear magnetic susceptibilities.  In a standard Pauli paramagnet, the linear splitting of the Fermi surface in applied magnetic fields does not affect the measured dHvA frequencies, but only their amplitudes through the spin-splitting effect~\cite{[20]}.  However, any non-linearity of the magnetisation as a function of magnetic field changes the situation considerably.  The main effects of relevance to this paper are that on the approach to metamagnetism, large changes are expected in the observed frequencies, and that in the high field state, magnetic splitting should be observable in the frequency spectrum~\cite{[21]}.  There is evidence for both in our data.  The field dependence of the main 1.8(6)~kT peak from the low (high) field states is shown in the inset to Fig. 2.  It is field-independent below 6.5~T, and above 10~T.  In the main part of Fig.~2 we show Fourier transforms of data from two regions (5.5~T $< B <~ $6.5~T and 10~T $ < B <~ $15~T) for which the observed frequencies are field-independent.  It is immediately clear that the high-field peaks have more structure, with evidence for splitting in each main peak.

Before discussing the metamagnetic change in the Fermi surface topography in more detail, we focus our attention on the low-field data.  Five main peaks are clearly resolved, with frequencies of approximately 0.2~kT, 0.5~kT, 0.8~kT, 1.8~kT and 4.2~kT respectively.  Initially, Sr$_3$Ru$_2$O$_7$ was reported to have the same body-centred tetragonal $I4mmm$ crystal structure as Sr$_2$RuO$_4$~\cite{[22]}, but later work~\cite{[23],[24]} identified $7^\circ$ rotations of the Ru-O octahedra, lowering the symmetry to $Bbcb$~\cite{[24]}. Mazin and Singh~\cite{[25]} have calculated the electronic band structure for both proposals of crystal structure, and have proposed that the distortion leads to major changes in the Fermi surface topography.  For the $I4mmm$ structure, the Fermi surface has the expected similarity with that of Sr$_2$RuO$_4$, with the small differences due to bilayer splitting, whereas the folding and extra gapping that occurs with the $Bbcb$ structure fragments the Fermi surface into many smaller portions.  Our data seem entirely consistent with the second scenario, since none of the frequencies that we observe below 2~kT would be predicted if the structure were $I4mmm$.  Moreover, the frequencies that we do observe fall qualitatively into the range of Fermi surface sheet areas predicted in the $Bbcb$ calculation.  Although the Fermi surface would be expected to reflect the true symmetry in this way, we note that in the perovskite CaVO$_3$, the observed Fermi surface did not reflect the gaps expected due to its orthorhombic distortion~\cite{[26]}.  The current observations indicate a strong electronic-structural coupling, since the induced gaps are evidently large enough to avoid magnetic breakdown in the range of fields studied here.

We have studied the effective masses of the quasiparticles on each orbit by fitting data taken between 0.1 K and 0.5 K to the standard Lifshitz-Kosevich formula.  The masses are fairly large: 6 $\pm$ 3, 13 $\pm$ 4, 10 $\pm$ 1, 8 $\pm$ 1 and 9 $\pm$ 2 electron masses for the 0.2~kT, 0.5~kT, 0.8~kT, 1.8~kT and 4.2~kT peaks respectively.  Since higher dHvA frequencies are damped out exponentially more strongly by impurity scattering, there is always some doubt about whether the entire Fermi surface including the largest sheets has been observed.  A standard check is to see whether the measured cyclotron masses can account for the specific heat.  That check is difficult to perform in the present case, because the Fermi surface gapping leads to the existence of multiple sheets each with the same area even in the first Brillouin zone.  The dHvA signal counts such `area degeneracies' only once, so a quantitative calculation of the Sommerfeld coefficient requires some assumptions.  The observed masses are large enough to be consistent with observation of the complete Fermi surface, but we cannot be definitive on this point.

%
\begin{figure}
\includegraphics[width=\columnwidth]{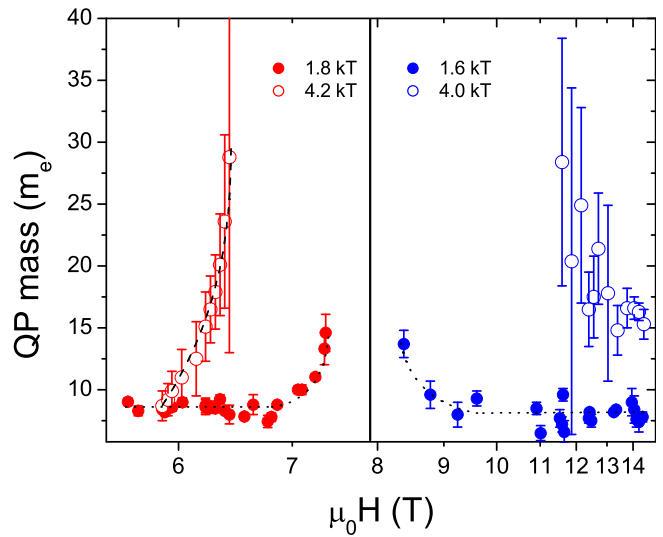}
\caption[]{(color online)Effective mass as a function of field for the peaks at 1.8(6)and 4.2(0) kT for the low (high) field states.  The dotted lines are guides to the eye.  The pronounced increase in the effective mass of the Landau quasiparticles can be understood in terms of fluctuations close to a quantum critical point.}
\end{figure}

The main interest of studying quantum oscillations in a metamagnetic system is not to calculate the low-field specific heat, but to obtain a microscopic picture of the behaviour of the Landau Fermi liquid above and below the transition~\cite{[16],[17],[18],[19]}.  One of the first steps in the analysis of the changes that take place at the transition is to identify what happens to each of the low-field Fermi surface sheets.  For this, we make use of the measured masses to deduce that the 0.2 and 0.5~kT peaks become the imperfectly resolved set of peaks below 0.4~kT, the 0.8~kT split peak probably becomes the pair of split peaks between 1.1 and 1.5~kT, the 1.8~kT peak shifts to the split pair whose prominent component is at 1.6~kT.  The 4.2~kT peak becomes the split peaks at approximately 4~kT.  

The splitting that we observe in the high-field state of all the main peaks is evidence that the small moment change of approximately $0.2 \mu_B/{\rm Ru}$~\cite{[10]} is due to exchange polarisation of the itinerant electrons in Sr$_3$Ru$_2$O$_7$.  In the simplest picture of itinerant metamagnetism, there is no need for a shift of the average (zero field) value of the dHvA peaks, but this ignores well-known consequences of itinerant metamagnetism such as structural changes due to magneto-structural coupling~\cite{[28]}.  The pressure coefficient of the metamagnetism in Sr$_3$Ru$_2$O$_7$ is known to be high~\cite{[29]}, and the low-field state is clearly sensitive to fairly fine structural detail, so it is no surprise to see overall changes in the observed frequencies.  These changes are rather small, suggesting that the basic 
picture appropriate to Sr$_3$Ru$_2$O$_7$ is much closer to `ideal exchange polarisation' than in other well-studied metamagnets such as UPt$_3$ and CeRu$_2$Si$_2$.  In UPt$_3$, the dHvA frequencies continue to be field-dependent even in regions where the magnetisation is field-linear~\cite{[16]}, which is beyond the scope of the simplest exchange polarisation analysis.  In CeRu$_2$Si$_2$, frequency changes through the metamagnetic transition are large, and the total Fermi surface volume is thought to change due to the localisation of the $f$ electrons~\cite{[17],[18],[19]}.  The present data do not give evidence for a substantial change in Fermi volume through the metamagnetic transition in Sr$_3$Ru$_2$O$_7$.  

The next issue that can be addressed experimentally using the quantum oscillation data is the field dependence of the quasiparticle effective mass.  Large changes in effective mass are expected in the vicinity of low temperature phase transitions, and have been seen in, for example CeRu$_2$Si$_2$~\cite{[17],[18],[19]}.  In Sr$_3$Ru$_2$O$_7$, a significant mass change has been inferred from the analysis of transport measurements~\cite{[3]} but it was clearly desirable to perform a direct thermodynamic measurement.  If we perform a field-dependent mass analysis in Sr$_3$Ru$_2$O$_7$, we resolve a mass change for the two sheets for which the signal 
can currently be tracked over an adequate range of temperature and field, as shown in the sample data set presented in Fig.~3.  In the quantum critical scenario discussed in detail in refs~\cite{[3],[10],[12]}, this mass change corresponds to renormalisation of the Landau quasiparticles due to the strong fluctuations in the vicinity of the quantum critical point.  The transport measurements \cite{[3]} pick up some fairly complicated FS average of the mass change, but dHvA measurements are capable of highlighting sheet-by-sheet differences.  A striking feature of Fig. 3 is the mass change associated with the 4 kT oscillation beginning much further from the metamagnetic field than that of the 1.6(8) kT oscillation, suggesting that this sheet is mainly responsible for the critical behaviour previously reported in ref. \cite{[3]}.

In summary, we have presented the results of a detailed study of quantum oscillations in Sr$_3$Ru$_2$O$_7$.  The observation and analysis of these oscillations has allowed the `coarse-grained' view of metamagnetic quantum criticality in this material to be studied from a microscopic rather than macroscopic view-point.  The results support a qualitative interpretation of the relevant physics as that of a field-induced Stoner transition whose critical fluctuations produce a significant renormalisation of the quasiparticle mass as the transition is approached.  The work also opens the way to future experiments that could be used to address some of the main outstanding issues.  To what extent, for example, is physics at high-$q$ responsible for the critical behaviour?  The data that we present on the topography of high- and low-field Fermi surface should be useful in 
interpreting inelastic neutron scattering experiments which have so far identified nesting-related scattering along the (1,0,0) direction on the low-field side.  Another question of central importance is how the quantum oscillations evolve and behave in the region between 7.5 and 8.3~T where they have not been resolved in the present experiment.  In future, it should be possible to improve our resolution by technical changes to the way we perform the dHvA experiments, identifying even purer crystals and possibly working in high hydrostatic pressure as was done on CeRu$_2$Si$_2$ in ref.~\cite{[19]}.  Successfully peforming those experiments would amount to a study of the fate of the Landau quasiparticle when subjected to divergent quantum fluctuations.

We are pleased to acknowledge discussions with C. Bergemann, S. R. Julian, G. G. Lonzarich and A. J. Millis.  This work was supported by the UK EPSRC, the Royal Society, the Leverhulme Trust, the 21COE program on ``Center for Diversity and
Universality in Physics'' from MEXT of Japan, by Grants-in-Aid for Scientific Research from Japan Society for Promotion of Science (JSPS), and MEXT.


\begin{thebibliography}{}
\bibitem{[14]} The data at the transition field still have a significant frequency dependent at this value of $\omega$.  See Ref.~\cite{[11]} for a discussion.
\bibitem{[21]} Both features are explained particularly clearly in  Fig. 1 of \cite{[16]}.
\end{thebibliography}
\end{document}